\documentclass[preprint2]{aastex}
\def\etal{\it et al.}
\def\eg{{\it e.g.,}\thinspace}
\def\ie{{\it i.e.,}\thinspace}

\shorttitle{}
\shortauthors{Basu, Mitra \& Rankin}

\begin{document}

\title{Toward an Empirical Theory of Pulsar Emission. X. On the Precursor and Postcursor Emission.}

\author{Rahul Basu$^{1,2}$, Dipanjan Mitra$^1$, Joanna M. Rankin$^3$}
\affil{$^1$National Centre for Radio Astrophysics, P. O. Bag 3, Pune University
Campus, Pune: 411 007. India.\\
$^2$ Kepler Institute of Astronomy, University of Zielona G\'{o}ra, Lubuska 2, 
65-265 Zielona G\'{o}ra, Poland. email:rahul@astro.ia.uz.zgora.pl\\
$^3$Department of Physics, University of Vermont, Burlington, VT 05401, USA.}

\begin{abstract}
\noindent
Precursors and postcursors (PPCs) are rare emission components detected in a 
handful of pulsars that appear beyond the main pulse emission, in some cases 
far away from it.  In this paper we attempt to characterize the PPC emission 
in relation to the pulsar main pulse geometry.  In our analysis we find that PPC 
components have properties very different from that of outer conal emission.  The 
separation of the PPC components from the main pulse center remains constant with 
frequency.  In addition the beam opening angles corresponding to the separation 
of PPC components from the pulsar center are much larger than the largest encountered 
in conal emission.  Pulsar radio emission is believed to originate within the magnetic 
polar flux tubes due to the growth of instabilities in the outflowing relativistic 
plasma.  Observationally, there is strong evidence that the main pulse emission 
originates at altitudes of about 50 neutron star radii for a canonical pulsar.  
Currently, the most plausible radio emission model that can explain main pulse emission 
is the coherent curvature radiation mechanism, wherein relativistic charged solitons 
are formed in a non-stationary electron-positron-pair plasma.  The wider 
beam opening angles of PPC require the emission to emanate from larger altitudes as 
compared to the main pulse, if both these components originate by the same emission 
mechanism. We explore this possibility and find that this emission mechanism is 
probably inapplicable at the height of the PPC emission.  We propose that the PPC emission 
represents a new type of radiation from pulsars with a mechanism different from that 
of the main pulse.
\end{abstract}


\section{\large Introduction}
\noindent
Pulsar emission is comprised of periodic pulses with varying shapes and 
intensities.  The integrated profile obtained by averaging a number of such 
pulses in phase (usually a few thousand pulses) is usually a stable feature 
showing little variation over time.  This pulsed emission usually has a high
degree of linear polarization with the polarization position angle (PPA) 
rotating across the profile with a characteristic S-shaped curve.  Such radio 
emission is thought to originate within the open magnetic field lines above 
the polar cap, centered around the magnetic axis, and is highly beamed.  
The pulsar profile consists of one or more features, known as components, 
which exhibit many different configurations and in certain cases frequency 
evolution.  The morphology and polarization properties of pulsar profiles are 
useful for determining the geometrical orientations of pulsars, the structure 
of their emission regions, their radio emission mechanisms, 
etc~\citep[hereafter ET~I, ET~IV]{lyn88,ran83,ran90}.

An empirical beam model for the main pulse radio emission has emerged
as a result of a wide array of studies conducted in the decades following the
discovery of pulsars~[ET~IV; \citet[hereafter ET~VIa]{ran93a}; \citet[hereafter 
ET~VII]{mit02}; \citet{mit99,gil00}].  The main pulse 
emission consists in general of a central core component surrounded by 
nested inner and outer cones of emission.  The rotating-vector model~\citep
{rad69}, which states that the plane containing the magnetic field lines and 
the rotation axis determines the orientation of the emitted radiation, accounts 
for the S-shaped swing of the PPA.  The core and the inner cone inhabit the 
regions of steeper PPA gradient, where the steepest gradient point (SG) is 
believed to fall near the longitude of the magnetic axis.  The outer conal 
emission occupies the profile wings with a relatively flat PPA.  The core and 
inner cones show nearly constant widths\footnote{The core reflecting the 
angular width of the polar cap at the stellar surface} and separations in most 
cases, whereas the outer conal emission exhibits radius-to-frequency mapping 
that shows increasing separation with decreasing frequency (ET~VII).  The 
polarization and spectral properties hint at two possibilities for the location of 
the outer and inner cones and core; either the different emission components
originate near the outer magnetic flux tube boundary in which case they are 
emitted at successively lower heights, or they arise from similar heights but 
occupy ever more central regions of the flux tube.  \citet{gg01} used aberration 
retardation effects to question the viability of the first possibility while a more 
theoretical approach to explaining the emission prefers the later one \citep{mac12}.  
The intensity of the various emission components varies, with one or more being 
weak or absent in certain pulsars and at high or low frequency, leading to 
asymmetric profiles known as ``partial cones'' [\citet{lyn88}; \citet[hereafter ET~IX]{mit11}.

The precursor and postcursor (PPC) emission found in a small sample of pulsars 
is noteworthy owing to their nonconformity with well understood conceptions of profile 
structure, such as the empirical core/double-cone model outlined above.  A precursor 
appears to be a distinct emission feature preceding the main pulse emission, whereas 
a postcursor follows it.  These PPC features are found further away from the main 
pulse center as indicated by the PPA traverse wherein the SG point falls within the 
main pulse.  The PPCs also exhibit a high degree of linear polarization and a relatively 
flat PPA, so resemble the outer conal emission.  However, unlike outer conal component 
pairs they are always asymmetric with respect to the main pulse profile center.  The 
presence of such PPCs has been reported in pulsars B0943+10 and B1822--09 by \citet{bac10} 
and also in B1322+83 and B2224+65 by ET~IX, where in each case they fall adjacent 
to a main pulse that is consistent with the above empirical model.  To our knowledge no 
previous study has been conducted to investigate the properties and geometry of the 
PPC emission features.  In this paper we endeavor to understand the location and 
physical properties of the PPC features within the context of the empirical beaming 
model developed in ET~VIa, \citet{mit99} and ETVII.  We have carried out a survey of the  
pulsar population to identify PPC features and studied their basic properties as a 
foundation for understanding them.

\S2 reviews the quantitative geometry of the empirical core/double-cone emission
beam model.  \S3 presents the results of our population survey wherein we attempt 
to distinguish PPC features from outer conal components.  In \S4 we investigate 
the implications of the PPC emission features from the perspective of their overall 
dipolar magnetic polar flux-tube geometry as a means of understanding their radio 
emission mechanism.

\section{\large Pulsar Emission Beam Structure} 
\noindent
Pulsar radio emission originates within the polar flux tube of open dipolar field lines 
above the magnetic poles.  The basic geometry of pulsar emission is determined by 
two angles: the so-called magnetic latitude angle between the rotation and magnetic 
axes $\alpha$ and the sightline impact angle between magnetic axis and observer's 
line of sight $\beta$. The rotating-vector model (RVM) describes the polarization 
position angle (PPA) $\chi$ at any longitude $\varphi$ in terms of this pulsar geometry 
as:
\begin{equation}
\small
{
\mbox{tan}\chi=\frac{\mbox{sin}\alpha~\mbox{sin}\varphi}{\mbox{cos}\alpha~\mbox{sin}(\alpha+\beta)-\mbox{sin}\alpha~\mbox{cos}(\alpha+\beta)~\mbox{cos}\varphi}
}.
\label{eqn_PPA}
\end{equation}
The steepest gradient point of the PPA, according to the RVM, is related to the 
geometry as:
\begin{equation}
\left(\frac{d\chi}{d\varphi}\right)_{\mbox{\tiny max}}=~~\frac{\mbox{sin}\alpha}{\mbox{sin}\beta}.
\label{eqn_SG}
\end{equation}
It is possible to solve for the latitude angle $\alpha$ and impact angle $\beta$ 
by carrying out RVM fits to the PPA traverse; however, the solutions are found to be highly 
correlated \citep{eve01}.  An independent means of estimating the latitude angle $\alpha$ was 
devised in ET~IV using the half power widths, $W_{\mbox{\small core}}$, of core emission 
components as \citep{mac11}: 
\begin{equation}
\mbox{sin}\alpha = 2.45\degr P^{-0.5}/W_{\mbox{\small core}}.
\label{eqn_corewid}
\end{equation}
The polar flux-tube radius angle, $\rho$, delimiting the emission region can be estimated 
using the overall pulse width, $W$, and the basic pulsar geometry \citep{gil84} as:
\begin{equation}
\small{
\mbox{sin}^2(\rho/2) = \mbox{sin}^2(\beta/2) + \mbox{sin}\alpha~\mbox{sin}(\alpha+\beta)~\mbox{sin}^2(W/4)
}.
\label{eqn_beamwid}
\end{equation}
A series of studies have been devoted to determining the beam opening angles of 
the conal emission, especially in pulsars with associated core emission, and 
ET~VIa,b discovered that the conal emission beams consist of a pair of 
inner and outer cones.  
\citet{mit99} analyzed around forty pulsars at six different frequencies and postulated 
the presence of three concentric conal beams with frequency evolution given as:
\begin{equation}
\rho^{\mbox{\footnotesize r}}_{\mbox{\tiny MHz}} = 4.8\degr \varrho(1 + (66\pm10)\nu^{-1\pm0.1}_{\mbox{\footnotesize MHz}})~\mbox{P}^{-0.5},
\label{eqn_conemit}
\end{equation}
with radii corresponding to $\varrho$ = 0.8, 1.0, 1.3 to within error of 0.03. 

The outermost conal emission seems to be associated with the last open magnetic 
field lines and thus provides direct estimates of the corresponding emission heights.
Taking into account divergence of the dipolar magnetic field lines with frequency, 
the emission height $r$ can be expressed in terms of the angular size of the polar cap
on the pulsar surface (radius 1.24\degr), the radius of the neutron star $R_S$ and the 
beam opening angle $\rho^{\mbox{\tiny outer}}$.
\begin{equation}
r = \mbox{P}\left(\frac{\rho^{\mbox{\tiny outer}}}{1.24\degr}\right)^2 R_S,
\label{eqn_hgeo}
\end{equation}
The main pulse emission based on the above geometrical analysis is estimated to arise 
at heights of about 20-50 $R_S$ for a typical pulsar [ET~VIa; \citet{kij98}; ET~VII].  
Emission-height estimates using aberration-retardation effects---\ie conal component 
shifts relative to the profile center as a result of relativistic beaming, which are 
independent of the pulsar geometry---are consistent with those estimated using the basic 
geometry above \citep{bla91,gg01,krz09}.

\section{\large PPC Emission and Wide Profiles}
\noindent
We have conducted an extensive population survey to identify pulsars with broad,
asymmetric profile features that might constitute PPC emission.  This also led to 
the identification of some additional pulsars with ``partial cones'' that appear to be 
extreme examples of outer conal emission (ET~IX).  We investigated pulsar 
profiles at multiple frequencies using various archival sources 
\citep{gou98,sei95,wei99,wei04} as well as unpublished observations from Arecibo 
and the GMRT.  The fourteen pulsars we identified with putative PPC emission are 
listed in Table \ref{tab_surv}.  The beam opening angles for the separation of 
candidate PPC features from the profile center were determined using two times the 
value of their separations from the main pulse centers.  The extremity of the outer conal 
emission (eq. \ref{eqn_conemit}) was used to distinguish between PPC and outer conal 
components (see Table \ref{tab_surv}).  For three pulsars, B0940+16, J1332--3032 and 
B1822--14, no published emission geometry was available; however, we were able 
to estimate parameter for one of them and so carry out the above analysis.  In the remainder 
of this section we describe each of the pulsars in more detail.

\begin{table*}
\caption{Emission geometry and beam opening angles at 1 GHz
\label{tab_surv}}
\begin{center}
\leavevmode
\begin{tabular}{lcccccccl}\hline\hline
  Pulsar  & Period &Profile& $\alpha$ & $\beta$ &Ref.& separation & $\rho$P$^{0.5}$& Comment \\
          &   (sec)  & Class &  ($\degr$) & ($\degr$) &&($\degr$)& ($\degr$)&         \\
\hline
  B0823+26   & 0.531 &{\bf S$_t$}     &  84~~  &  +1.9 & 1 & ~41.6 &  30.2  &postcursor \\
  B0940+16   & 1.087 &{\bf D}?        &$\sim$32&$\sim$0&   & ~85~~ &$\sim$44&postcursor?\\ 
  B0943+10   & 1.098 &{\bf S$_d$}     &  11.6  &  ~4.31& 2 & ~52.3 &  13.2  &precursor  \\
  B0950+08   & 0.253 &{\bf S$_d$}     &  16~~  &--11.3 &   & 150~~ &  10.2  &precursor  \\
  B1322+83   & 0.670 &{\bf S$_d$}     &  14~~  &  ~5.1 & 3 & ~42.6 &  10.5  &precursor  \\
  J1332--3032& 0.650 &     ?          &   ---  &  ---  &   & 119.0 &  ---   &precursor? \\ 
  B1524--39  & 2.418 &  {\bf D}       &  16.5  & --0.4 & 5 & ~57.7 &  12.5  &precursor  \\ 
  B1530+27   & 1.125 &{\bf S$_d$/D}   &  30~~  &  +4.9 & 3 & ~52~~ &  29.2  &postcursor \\ 
  B1742--30  & 0.367 &{\bf T}         &  24~~  &  ~6.4 & 3 & ~14.3 &  ~5.5  &outer cone \\ 
  B1822--09  & 0.769 &{\bf T}         &  86~~  &  ~0.0 & 4 & ~14.3 &  12.5  &precursor  \\ 
  B1822--14  & 0.279 &           ?    &   ---  &  ---  &   & ~26.4 &   ---  &precursor? \\ 
  B1929+10   & 0.227 &{\bf T/M}?      &  90    &  41.8 &   & 115~~ &  51.6  &postcursor \\ 
  B2217+47   & 0.538 &{\bf S$_t$}     &  42~~  &  ~4.5 & 3 & 42--68& 23--35 &postcursor \\ 
  B2224+65   & 0.683 &{\bf T$_{1/2}$} &  15.2  &  ~3.3 & 5 & ~30.1 &  ~7.6  &outer cone \\ 
\tableline
\end{tabular}
\tablenotetext{~}{1. ET~VIb; 2. \citet{ran03}; 3. ET~IX; 4. \citet{bac10}; 5. this paper}
\end{center}
\end{table*}

\begin{figure*}
\begin{tabular}{@{}lr@{}}
{\mbox{\includegraphics[angle=0,scale=0.6]{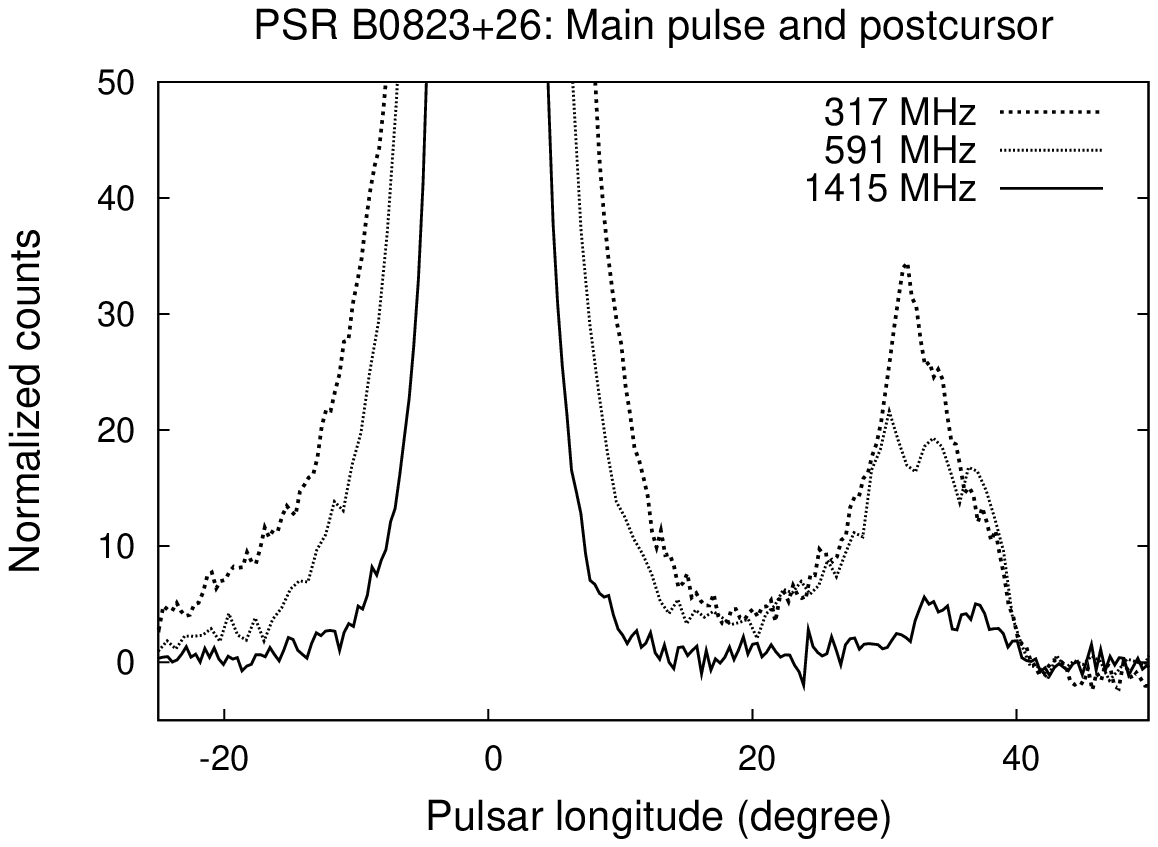}}} &
{\mbox{\includegraphics[angle=0,scale=0.6]{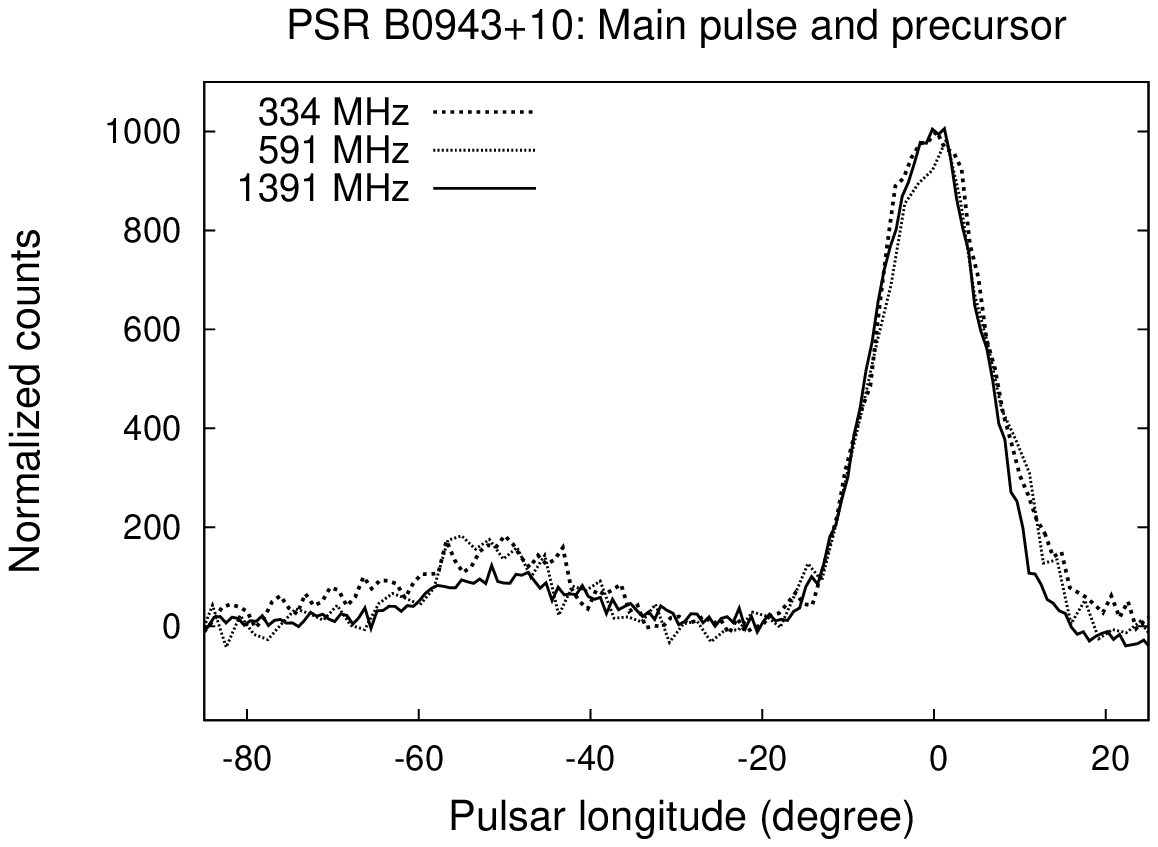}}} \\
{\mbox{\includegraphics[angle=0,scale=0.6]{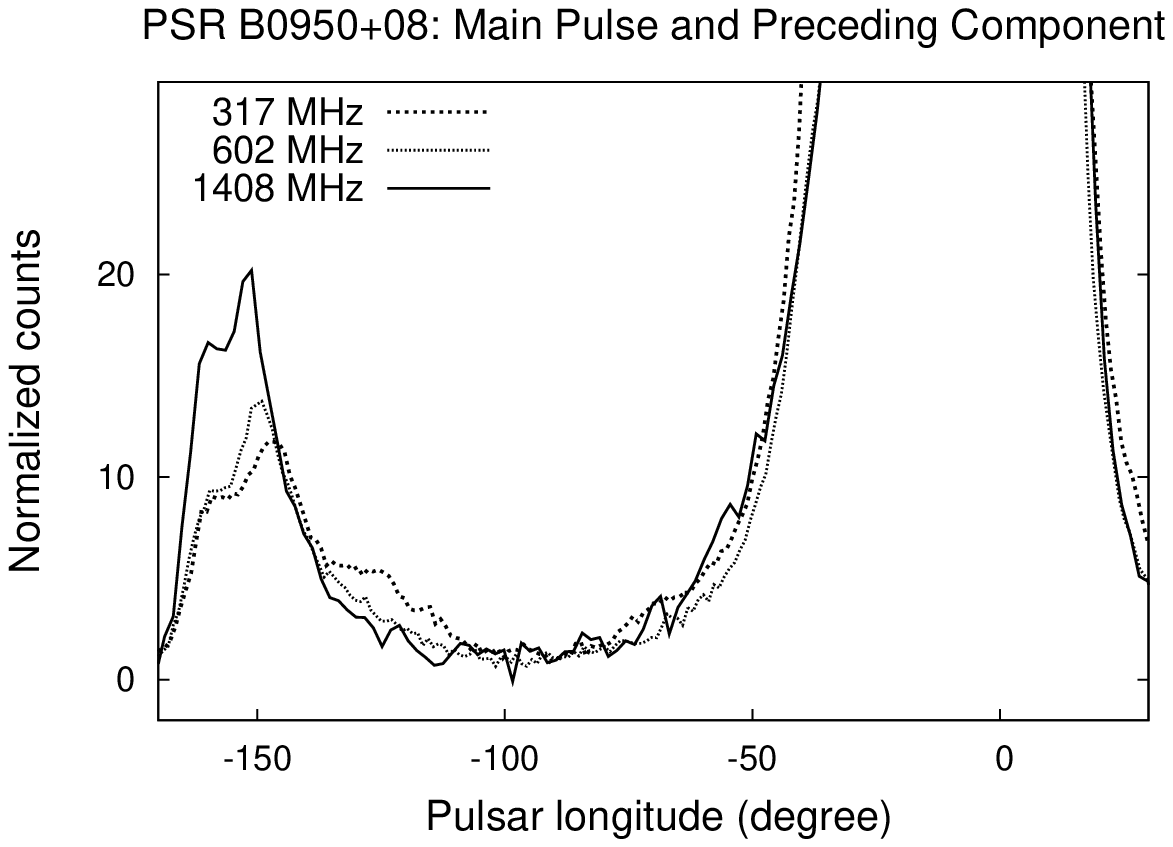}}} &
{\mbox{\includegraphics[angle=0,scale=0.6]{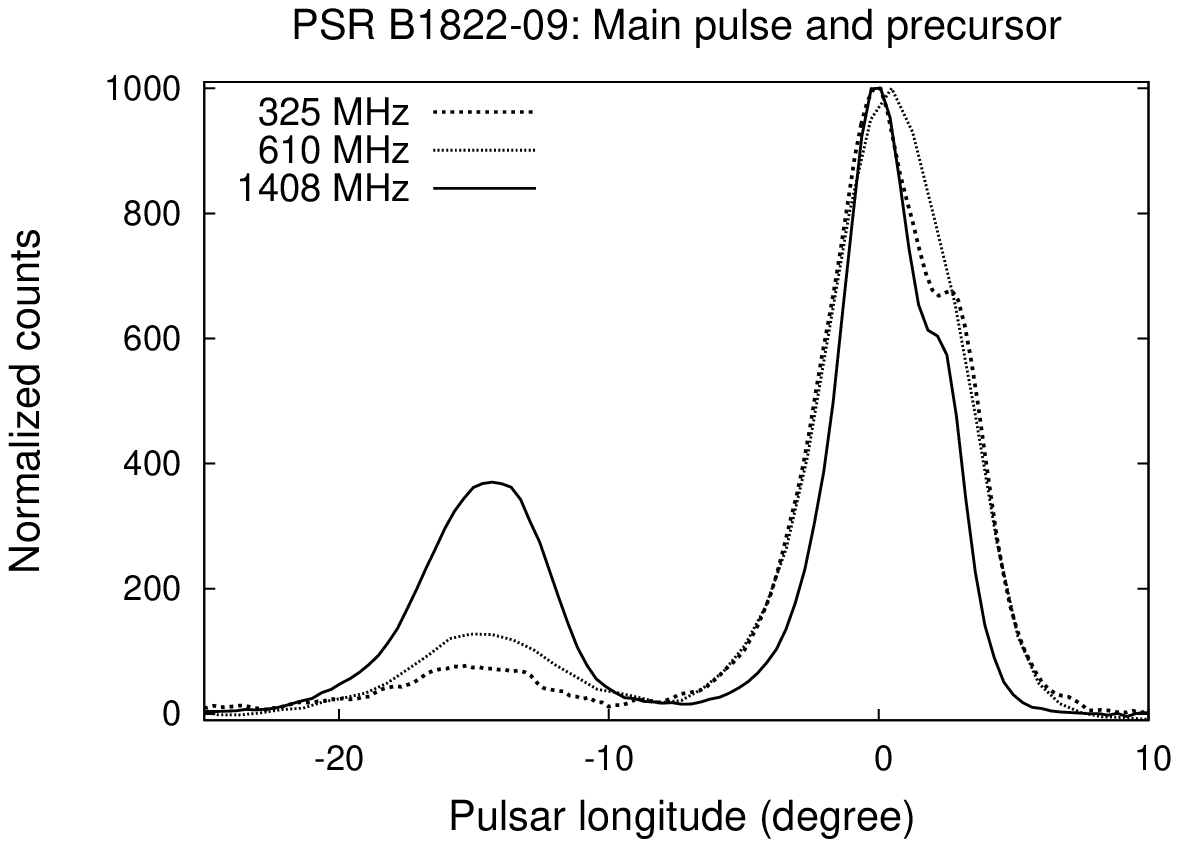}}} \\
{\mbox{\includegraphics[angle=0,scale=0.6]{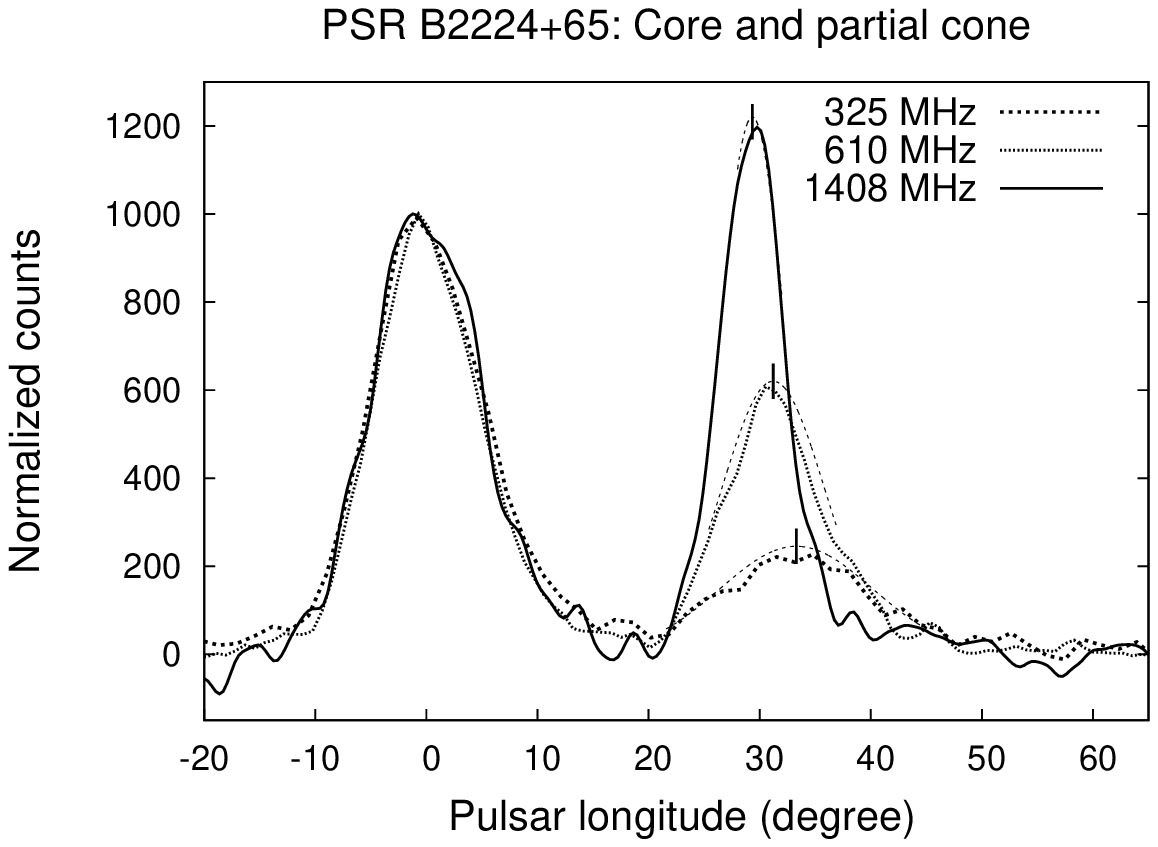}}} &
{\mbox{}} \\
\end{tabular}
\caption{\small Frequency evolution of the postcursor in pulsar B0823+26 (top left), the 
precursor in B0943+10 (top right), the precursor in B0950+08 (middle left), the precursor 
in B1822--09 (middle right) and the ostensible partial cone in B2224+65 (bottom left). The precursors 
in B0943+10 and B1822--09 maintain a constant separation from the main pulse center with 
frequency. The precursor is brighter at higher frequencies in the case of B1822--09. The 
precursor remains more or less constant in B0943+10~in our frequency range, however, it is 
known to get considerably weaker at lower frequencies. The precursor in B0950+08 shows a
slight evolution with frequency where the emission grows stronger and moves further away 
from the main pulse with increasing frequency.  However, the main pulse in this pulsar also
shows significant frequency evolution, and the perceived shift may be a result of erroneous
determination of pulse center at each frequency. The postcursor in B0823+26 consists of 
three subfeatures, all 
of which grow weaker with increasing frequency.  The leading feature is most prominent 
at 317 MHz and is barely detected at 1.4 GHz.  However, the outer edge of the postcursor emission 
maintains a constant separation from the main pulse.  The outer conal emission in B2224+65 which 
here is identified as a partial cone becomes more prominent with increasing frequency relative to 
the central core component (the latter typically have steeper RF spectra).  The separation of the 
conal component from the pulse center as well as the component width decreases with increasing frequency.  
\label{fig_comp}}
\end{figure*}

\noindent
{\bf B0823+26} exhibits a well established postcursor feature in addition to its main 
pulse and interpulse \citep{han86}.  \citet{ran95} studied the pulsar's profile regions 
in detail and appeared to confirm an increasing postcursor separation with frequency.  
The basic geometry was modeled in ET~VIb and the values reported in Table 
\ref{tab_surv}.  Profiles for this pulsar including its postcursor have been measured 
over the frequency range from 317 MHz to 1.4 GHz and their various characteristics given 
in Table \ref{tab_morph}.  The interpulse to main pulse separation is nearly 180$\degr$ 
over this band, strongly suggesting an orthogonal geometry.  However, we find that the 
postcursor is comprised of three features which have different spectra as shown in 
Figure~\ref{fig_comp}.  The leading feature is prominent at lower frequencies but hardly 
detectable above 1 GHz---thus use of the postcursor peak spacing leads to flawed conclusions 
about its frequency evolution.  The trailing extremity of the postcursor provides a better 
estimator of its separation.  In Table~\ref{tab_morph} we list the half widths of the 
main pulse, the separations of the postcursor feature (5-$\sigma$ at the outer edge) from 
the pulse center, and the beam opening angles for  both main pulse and the postcursor 
separation at all available frequencies.  Therefore, the postcursor separation does 
not appear to vary with frequency contrary to \citet{han86} and \citet{ran95}. The frequency 
evolution of the postcursor emission beam and the corresponding emission height using 
eq.(\ref{eqn_hgeo}) is shown in Figure~\ref{fig_rad}.  

\noindent
{\bf B0940+16} shows what appears to be a postcursor, at 430 MHz, about 85$\degr$ 
from the center of its main pulse \citep{dei86,han10}.  The feature is also present at 1.4 GHz 
\citep{wei99}, though it is much weaker.  The PPA shows a relatively flat swing across 
the main pulse suggesting conal emission cut centrally by the sightline.  We then 
tentatively class this pulsar as exhibiting an inner cone double profile with a $\beta$ 
value near 0\degr, such that $\alpha$ would then be about 32\degr. The beam opening angle
(normalized by the pulsar period) is around 44\degr (see Table~\ref{tab_surv}) which lies
outside the conal boundary.

\noindent
{\bf B0943+10} is well known for its so-called quiescent (Q) and burst (B) modes, and 
its profile evolution and highly regular subpulse drifting in the latter establish it as a 
classic example of conal emission.  \citet{bac10} discovered a profile feature leading
the main pulse during only the Q mode that they identified as a precursor.  We had 
access to Q-mode profiles at 325, 610 and 1391 MHz which were used for our investigation 
here.  The pulsar is weaker during the Q mode, and adequately sensitive profiles at 
higher frequencies are difficult to obtain owing to its steep RF spectrum.  ET~VIb
computed its basic geometry, but more accurate values come from more recent models 
of its subbeam carousel \citep{des01,ran03}, and these are the values reported in 
Table~\ref{tab_surv}.  We have determined the width of the main pulse, the separation 
of the precursor from the pulse center and the beam opening angles as shown in Table
\ref{tab_morph}. 

\noindent
{\bf B0950+08} shows the presence of a leading component about 150\degr~from the 
main pulse.  This leading component has usually been considered an interpulse, and
questions about whether the combination represented a nearly aligned or orthogonal 
rotator (one- or two-pole configuration) much debated [\citet{bla91}; ET~VIb; \citet{eve01}].  On 
its own, however, the evidence is strong that the main pulse represents an outer conal single 
geometry given the prominent bifurcation of its profile at low frequencies \eg \cite{han10}.
Moreover, the very shallow PPA rate and strong indication of a 360\degr\ traverse over 
each rotation (\eg \citet{bla91}) almost definitively signals a small $\alpha$ and negative 
$\beta$ similar to the calculations in ET~VIa,b.  The main pulse is joined to the 
preceding component via a bridge emission, which also shows high degree of linear polarization 
and is located far away from the SG (located close to the main pulse peak) suggesting a 
possible precursor emission. We used high quality GMRT and Arecibo data with frequencies 
ranging from 111 MHz to 2370 MHz, with the main pulse showing considerable profile evolution 
across frequencies. As a result it was difficult to identify the central point and we used 
the mid point of the half widths as reference for determining the separation between components.
The separation between the two components shows a slight frequency evolution and increases 
with increasing frequency (see Figure~\ref{fig_comp} and Table \ref{tab_morph}) which may be due 
to errors in determining the pulse centre. We used the outer conal geometry as estimated by 
ET~VIb [with correction of sign in $\beta$ following the convention of \cite{eve01}] 
with values of $\alpha$ = 16\degr\ and $\beta$ = --11.3\degr. The period normalized beam width 
for the leading emission component at 1 GHz turns out to be 10\degr\ as seen in other precursors. 
The leading emission in this pulsar is then most likely a precursor but may show a slight 
frequency evolution.

\noindent
{\bf B1322+83} was studied in detail in ET~IX at 325 MHz, who interpreted its 
profile as having a precursor emission feature.  The main pulse emission exhibits the 
properties of a conal single profile similar to the Q mode of B0943+10. The pulsar also 
shows signatures of interstellar scattering which made estimating the feature widths 
difficult.  The geometry of the pulsar was determined in ET~IX (see Table~\ref{tab_surv}).  
The precursor feature is particularly weak in this pulsar and was only detected at one 
other frequency, 610 MHz.  We were unable to estimate the width of the precursor 
at 610 MHz due to its weakness, though its peak could be identified. The main pulse
and precursor widths and their separation at 325 and 610 MHz are shown in 
Table~\ref{tab_morph} along with the radius of the beam opening angles.  

\noindent
{\bf J1332--3032} The pulsar exhibits a two component profile at low 
frequencies $<$ 1 GHz (325, 430 and 610 MHz). The leading component is 
119$\degr$ ahead of the main pulse emission with the separation remaining 
constant across frequencies and is therefore ostensible precursor emission. 
The pulsar is relatively weak and lacks polarization measurements at any of 
the frequencies. We do not have any estimate of the pulsar's geometry and 
this makes it difficult to distinguish between a wider inner cone and precursor 
emission.

\noindent
{\bf B1524--39} appears to have an emission feature about 58$\degr$ ahead of 
its main pulse.  This putative precursor emission was detected only at 435 MHz 
and was not detected at 325 and 610 MHz.  The main pulse profile has a conal 
double form with a steep 180$\degr$ PPA sweep across it and shows little or no 
width evolution with frequency.  The geometry of the pulsar was determined by 
fitting the RVM to the PPA at 325 MHz and is shown in Table \ref{tab_surv}.  The 
beam opening angle for the separation of the precursor feature from the 
profile center was well outside the conal range (see Table \ref{tab_surv}).  We 
note the peculiarity of the precursor detection at a single frequency and its 
absence at both 325 and 610 MHz.  This suggests the possibility that the pulsar 
has modes like those of B0943+10 wherein precursor emission is present only 
during a certain emission mode.  The pulsar needs to be observed for longer 
durations to assess whether mode changing could be an issue.

\noindent
{\bf B1530+27} shows weak putative postcursor emission 52$\degr$ from its main 
pulse peak.  The pulsar has been studied in detail at 325 MHz in ET~IX
where its main pulse was identified as conal.  Its main pulse shows a double form 
and a shallow PPA traverse implying a probable oblique sightline cut through the 
emission beam.  The above study also determined the pulsar's geometry which 
is given in Table~\ref{tab_surv}.  The postcursor feature is also seen at 430 MHz  
\citep{wei04} at the same separation from the main pulse.  The beam opening 
angle for the separation of postcursor emission from the pulse center is 
shown in Table \ref{tab_surv} which clearly lies outside the conal range.

\noindent
{\bf B1742--30} A leading component located roughly 15$\degr$ ahead of the 
main pulse is seen in this pulsar over a wide frequency range.  The separation 
of this feature from the trailing components increases with wavelength.  A low 
level counterpart trailing the central component is also visible in certain higher 
sensitivity profiles.  The geometry of the pulsar was determined in ET~IX and 
is shown in Table \ref{tab_surv}.  The beam opening angle at 1 GHz for the 
separation of the leading component from the pulse center is calculated 
(see Table \ref{tab_surv}) and lies within the conal region as per eq.(\ref{eqn_conemit}).  
The leading emission component is thus identified as outer conal in this 
pulsar which exhibits such an unusually broad triple ({\bf T}) profile.

\noindent
{\bf B1822--09}'s profile also famously exhibits both a leading precursor and an 
interpulse in addition to its main pulse, and these emission features are seen 
alternately in its quiescent (Q) and burst (B) modes. The interpulse appears in 
its Q mode and the precursor in the B.  However, there are mixed-mode episodes 
where both emission components are present (or absent) simultaneously \citep{lat12}. 
\citet{gil94} studied this pulsar extensively at several frequencies, interestingly 
reporting microstructures in the leading feature, but failed to identify it as having 
the unique properties of a precursor.  \citet{bac10} first established the core/inner 
cone character of the pulsar's main pulse in relation to the leading precursor 
feature.  The geometry of this pulsar has been computed in ET~IX and given in 
Table \ref{tab_surv}.  We studied the pulsar over a wide frequency range from 
240 MHz to 10.5 GHz with the various component widths and beam opening 
radii summarized in Table~\ref{tab_morph}.  The precursor emission becomes 
relatively weaker with decreasing frequencies (see Figure~\ref{fig_comp}).  
The beam opening angle for the separation of precursor emission from pulse 
center lies well outside the conal region and does not evolve with frequency 
(see Figure~\ref{fig_rad}).

\noindent
{\bf B1822--14}  shows an emission feature preceding its main pulse, faintly at 
1.4 - 1.6 GHz and imperceptible below \citep{gou98} but quite brightly at 5 GHz 
\citep{hoe99}.  No published geometry for the pulsar is available and the evidence 
is limited for us to characterize the main pulse as well as the leading component.

\noindent
{\bf B1929+10}'s  profile consists of a core/(double?)cone main pulse and a 
core-single interpulse.  As discussed in ET~VIa,b, the widths of both cores indicate 
an orthogonal (two-pole) geometry.   The main pulse and interpulse are separated 
by ``bridges'' of weak highly linearly polarized emission over essentially the entire 
rotation cycle; however, a somewhat stronger postcursor feature follows the main 
pulse.  The separation between the main pulse and interpulse is 187.4\degr\ and 
remains constant with frequency~\citep{han86}.  At odds with this view are RVM 
fits to the ``bridge'' emission [that necessarily ignore the main pulse; \cite{eve01}] 
suggesting $\alpha$ and $\beta$ values of some 36\degr and 26\degr.  Given that these 
values permit no sense to be made of the obsensible {\bf T} or {\bf M} main-pulse 
component structure, we adopt the above two-pole geometry for our purposes here 
(and wonder if the low level ``bridge'' emission has a different source).  The pulsar's 
off-pulse emission has been studied in detail by \citet{ran97}.  The putative postcursor 
component is fully linearly polarized, follows the main-pulse peak by around 115\degr, 
and its position seems to remain constant with frequency. 

\noindent
{\bf B2217+47} The presence of a postcursor component in this pulsar at 102.5 
MHz was reported by \citet{sul94}. The postcursor was connected to the main 
pulse via an emission bridge and was distinguished by a gradual time evolution.  
Its intensity increased with time (during the observing cycle lasting several years), 
and it moved towards the main pulse resulting in a decreasing separation between 
the components.  The temporal variations of the postcursor were attributed to 
precession of the rotation axis.  The postcursor is absent in higher frequencies 
despite the presence of good quality profiles.  The geometry of the pulsar has 
been determined in ET~IX with the $\alpha$ and $\beta$ values shown in 
Table~\ref{tab_surv}.  The main pulse is classified as core-single.  The corresponding 
beam opening angle for the separation of the postcursor from pulse center 
has been determined in Table~\ref{tab_surv} and turns out to be large like other 
postcursor emission.

\noindent
{\bf B2224+65} shows a profile with two widely separated emission components, 
and several studies have interpreted these differently:  \citet{lyn88} classified the 
profile as being a ``partial cone'' while ET~IX identified the trailing feature as 
a postcursor in relation to a core-single main pulse.  The latter study estimated the 
geometry of the pulsar as having an $\alpha$ of 27$\degr$ and $\beta$ of 4.9$\degr$. 
The core component width (see Table~\ref{tab_morph} and eq. \ref{eqn_corewid}) 
was incompatible with the above geometry.  Using our measured core widths we 
established the corrected geometry for this pulsar to be $\alpha$ = 15.2$\degr$ 
and $\beta$ = 3.3$\degr$.\footnote{We used $\left(d\chi/d\varphi\right)_{\mbox{\tiny max}}$ =
-4.5 from \citet{lyn88} in our calculations of $\beta$ for the pulsar B2224+65.} The 
pulsar was studied over frequencies ranging from 325 MHz to 1.6 GHz and the 
various component widths and beam opening angles are shown in Table~\ref{tab_morph}.  
The trailing component grows stronger and its separation decreases with increasing 
frequency (see Figure~\ref{fig_comp}).  In addition the beam opening angles lie on 
the border of outer conal beam as seen in Figure~\ref{fig_rad}.  We identify the trailing 
component to be outer conal emission, a classic example of partial cone, lying on the 
extreme edge of the emission beam. 

\begin{table*}
\caption{The component spacing, width and beam radius}
\label{tab_morph}
\begin{center}
\leavevmode
\begin{tabular}{cccccccc}\hline\hline
 Pulsar & Period & $\nu$ & MP width & PC width & PC-MP sep & $\rho^{\mbox{\tiny MP}}$ & $\rho^{\mbox{\tiny PC}}$ \\
  & (sec)& (MHz) & ($\degr$) &($\degr$)&($\degr$)&($\degr$)&($\degr$)\\
\hline
          &       & ~317 & 5.4$\pm$0.1 &      ---      & 41.4$\pm$0.4 & 3.3$\pm$0.04& 41.3$\pm$0.1 \\
          &       & ~408 & 4.1$\pm$0.1 &      ---      & 42.8$\pm$0.8 & 2.8$\pm$0.04& 42.7$\pm$0.2 \\
          &       & ~430 & 4.1$\pm$0.1 &      ---      & 42.0$\pm$0.4 & 2.8$\pm$0.04& 41.9$\pm$0.1 \\
 B0823+26 & 0.531 & ~591 & 5.0$\pm$0.1 &      ---      & 41.7$\pm$0.7 & 3.1$\pm$0.04& 41.6$\pm$0.2 \\
          &       & ~610 & 4.1$\pm$0.1 &      ---      & 41.2$\pm$0.7 & 2.8$\pm$0.04& 41.1$\pm$0.2 \\
          &       & 1178 & 2.7$\pm$0.1 &      ---      & 41.4$\pm$0.2 & 2.3$\pm$0.03& 41.3$\pm$0.1 \\
          &       & 1404 & 3.2$\pm$0.1 &      ---      & 42.1$\pm$0.7 & 2.5$\pm$0.03& 42.0$\pm$0.2 \\
          &       & 1415 & 3.4$\pm$0.1 &      ---      & 41.8$\pm$0.4 & 2.5$\pm$0.03& 41.7$\pm$0.1 \\
\hline
          &       & ~325 & 12.8$\pm$0.1 & 23.4$\pm$1.0 & 52.8$\pm$0.6 & 4.6$\pm$0.05 & 12.7$\pm$0.1 \\
 B0943+10 & 1.098 & ~610 & 13.2$\pm$0.2 & 14.4$\pm$1.2 & 51.7$\pm$0.7 & 4.6$\pm$0.05 & 12.5$\pm$0.1 \\
          &       & 1391 & 15.2$\pm$0.7 & 18.7$\pm$1.0 & 50.2$\pm$0.8 & 4.7$\pm$0.05 & 12.3$\pm$0.1 \\
\hline
          &       & ~111 & 19.7$\pm$0.4 &      ---     & 144.9$\pm$0.9 & 11.4$\pm$0.05 & 20.0$\pm$0.1\\
          &       & ~317 & 14.7$\pm$0.4 & 25.2$\pm$1.0 & 146.3$\pm$0.4 & 11.4$\pm$0.05 & 20.1$\pm$0.1\\
 B0950+08 & 0.253 & ~430 & 14.1$\pm$0.4 & 23.2$\pm$1.1 & 147.3$\pm$0.5 & 11.3$\pm$0.05 & 20.1$\pm$0.1\\
          &       & ~602 & 12.3$\pm$0.3 & 22.0$\pm$1.0 & 149.1$\pm$0.4 & 11.3$\pm$0.05 & 20.2$\pm$0.1\\
          &       & 1408 & 14.4$\pm$0.4 & 15.8$\pm$1.8 & 151.2$\pm$0.4 & 11.4$\pm$0.05 & 20.2$\pm$0.1\\
          &       & 2307 & 12.7$\pm$0.4 & 16.9$\pm$1.1 & 153.5$\pm$0.8 & 11.3$\pm$0.05 & 20.3$\pm$0.1\\
\hline
 B1322+83 & 0.670 & ~325 & ~9.8$\pm$0.2 &  17.1$\pm$1.0 & 42.9$\pm$0.6 & 5.3$\pm$0.1 & 12.9$\pm$0.1 \\
          &       & ~610 & 10.0$\pm$0.4 &      ---      & 42.2$\pm$0.8 & 5.3$\pm$0.1 & 12.7$\pm$0.2 \\
\hline
           &       & ~~243 & 8.1$\pm$0.4 & 3.9$\pm$0.6 & 14.4$\pm$0.3~& 4.0$\pm$0.2 & 14.4$\pm$0.3 \\
           &       & ~~325 & 7.1$\pm$0.2 & 7.0$\pm$0.2 & 15.2$\pm$0.1~& 3.5$\pm$0.1 & 15.2$\pm$0.1 \\
           &       & ~~408 & 6.7$\pm$0.4 & 4.1$\pm$0.2 & 14.8$\pm$0.1~& 3.3$\pm$0.2 & 14.8$\pm$0.1 \\
           &       & ~~610 & 7.1$\pm$1.1 & 6.7$\pm$0.2 & 14.6$\pm$0.6~& 3.5$\pm$0.5 & 14.6$\pm$0.6 \\
           &       & ~~800 & 6.8$\pm$0.1 & 7.4$\pm$0.3 & 14.5$\pm$0.1~& 3.4$\pm$0.1 & 14.5$\pm$0.1 \\
           &       & ~~925 & 5.4$\pm$0.1 & 7.7$\pm$0.2 & 15.0$\pm$0.1~& 2.7$\pm$0.1 & 15.0$\pm$0.1 \\
           &       & ~1330 & 6.5$\pm$0.1 & 7.1$\pm$0.1 & 14.7$\pm$0.1~& 3.2$\pm$0.1 & 14.7$\pm$0.1 \\
 B1822--09 & 0.769 & ~1408 & 5.2$\pm$0.1 & 6.0$\pm$0.1 & 14.7$\pm$0.1~& 2.6$\pm$0.1 & 14.7$\pm$0.1 \\
           &       & ~1408 & 5.5$\pm$0.2 & 5.4$\pm$0.1 & 14.6$\pm$0.1~& 2.7$\pm$0.1 & 14.6$\pm$0.1 \\
           &       & ~1410 & 6.1$\pm$0.3 & 5.6$\pm$0.2 & 13.6$\pm$0.1~& 3.0$\pm$0.2 & 13.6$\pm$0.1 \\
           &       & ~1640 & 5.7$\pm$0.3 & 6.4$\pm$1.0 & 14.9$\pm$0.4~& 2.8$\pm$0.2 & 14.9$\pm$0.4 \\
           &       & ~1642 & 6.0$\pm$1.1 & 5.6$\pm$0.1 & 14.3$\pm$0.8~& 3.0$\pm$0.5 & 14.3$\pm$0.8 \\
           &       & ~4750 & 5.9$\pm$0.7 & 5.1$\pm$0.1 & 14.1$\pm$0.4~& 2.9$\pm$0.3 & 14.1$\pm$0.4 \\
           &       & ~4850 & 4.9$\pm$0.1 & 5.6$\pm$0.1 &~13.9$\pm$0.04& 2.4$\pm$0.1 & 13.9$\pm$0.04\\
           &       & 10450 & 3.0$\pm$0.2 & 5.5$\pm$0.5 & 14.3$\pm$0.2~& 1.5$\pm$0.1 & 14.3$\pm$0.2 \\
           &       & 10550 & 3.6$\pm$0.1 & 6.3$\pm$0.3 & 14.3$\pm$0.1~& 1.8$\pm$0.1 & 14.3$\pm$0.1 \\
\hline
          &       & ~325 & 12.1$\pm$0.2 & 18.7$\pm$0.7 & 33.3$\pm$0.3 & 3.7$\pm$0.2 & 10.0$\pm$0.1 \\
          &       & ~400 & 12.0$\pm$0.2 & 15.1$\pm$0.7 & 32.1$\pm$0.3 & 3.7$\pm$0.2 & ~9.7$\pm$0.1 \\
          &       & ~408 & 10.4$\pm$0.3 & 13.4$\pm$1.3 & 32.7$\pm$0.6 & 3.6$\pm$0.2 & ~9.9$\pm$0.2 \\
          &       & ~610 & 11.3$\pm$0.2 & 10.9$\pm$0.3 & 31.2$\pm$0.1 & 3.7$\pm$0.2 & ~9.5$\pm$0.1 \\
 B2224+65 & 0.683 & ~800 & 12.0$\pm$0.5 & 10.1$\pm$0.4 & 30.5$\pm$0.3 & 3.7$\pm$0.2 & ~9.3$\pm$0.1 \\
          &       & ~925 & 11.3$\pm$0.5 & ~7.5$\pm$0.3 & 29.9$\pm$0.2 & 3.7$\pm$0.2 & ~9.2$\pm$0.1 \\
          &       & 1330 & 12.3$\pm$0.3 & ~8.3$\pm$0.4 & 29.5$\pm$0.2 & 3.7$\pm$0.2 & ~9.0$\pm$0.1 \\
          &       & 1408 & 10.7$\pm$0.1 & ~6.7$\pm$0.1 & 29.5$\pm$0.2 & 3.6$\pm$0.2 & ~9.0$\pm$0.1 \\
          &       & 1408 & 11.7$\pm$0.2 & ~7.0$\pm$0.1 & 29.3$\pm$0.1 & 3.7$\pm$0.2 & ~9.0$\pm$0.1 \\
          &       & 1642 & 11.5$\pm$0.3 & ~6.7$\pm$0.1 & 28.9$\pm$0.1 & 3.7$\pm$0.2 & ~8.9$\pm$0.1 \\
\hline
\end{tabular}
\end{center}
\end{table*}

\begin{figure*}
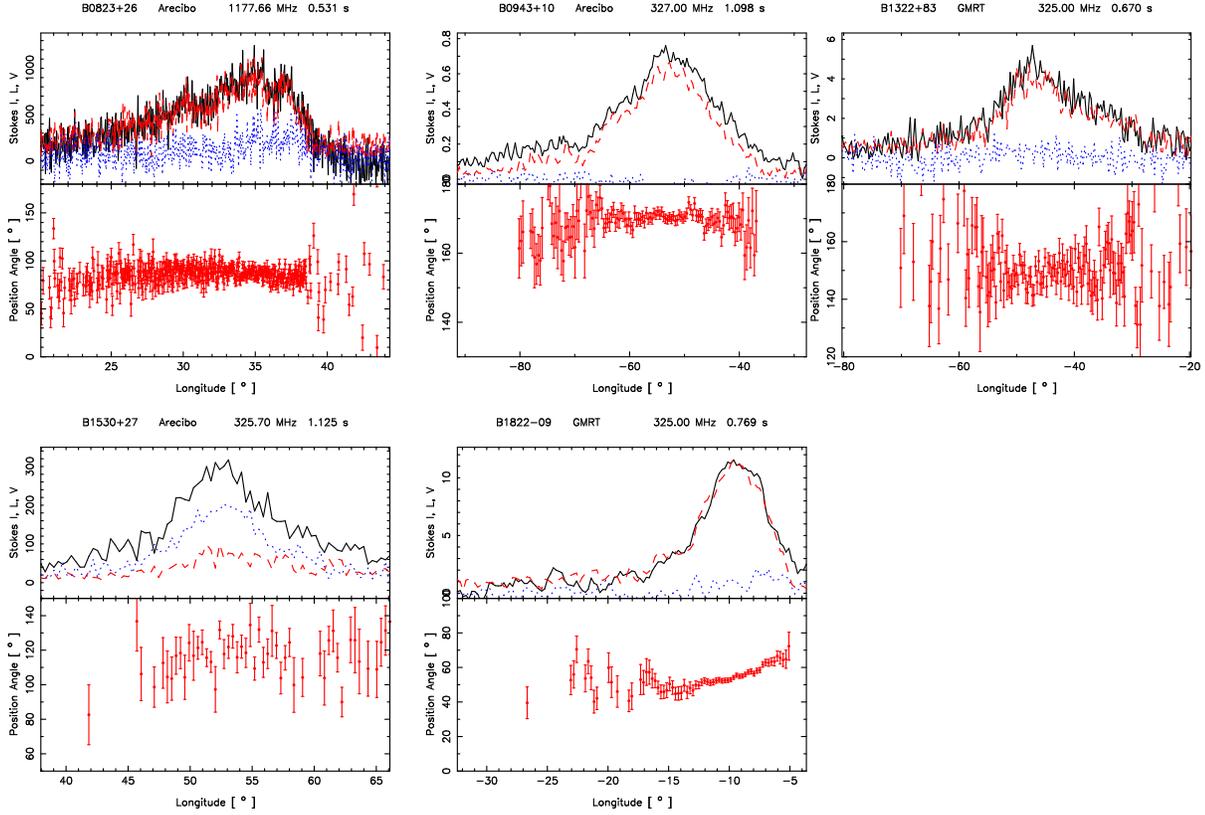

\begin{tabular}{@{}lr@{}lr@{}}
{\mbox{\includegraphics[angle=-90,scale=0.29]{B0823+26_1400_pc.ps}}} &
{\mbox{\includegraphics[angle=-90,scale=0.29]{B0943+10_1400_pc.ps}}} &
{\mbox{\includegraphics[angle=-90,scale=0.29]{B1322+83_325_pc.ps}}} \\
{\mbox{\includegraphics[angle=-90,scale=0.29]{B1530+27_325_pc.ps}}} &
{\mbox{\includegraphics[angle=-90,scale=0.29]{B1822-09_325_pc.ps}}} &
{\mbox{}} \\
\end{tabular}
\caption{Polarization characteristics of pulsars with the PPC emission: total 
intensity I (black line), the linear polarization L (red broken line), circular polarization V 
(blue dotted line) as well as the polarization position angle (PPA).  The postcursor 
in pulsar B0823+26 (top left) shows high linear polarization ($>$ 70 \%) and 
moderate circular polarization ($<$ 40\%).The PPA shows no discernible swing 
across the postcursor feature.  In B0943+10 (top center) the precursor has high linear 
polarization and low circular polarization with the PPA remaining flat across the 
precursor.  In the case of the precursor in B1322+83 (top right), once again we see 
high linear polarization ($>$ 75\%) but low circular polarization ($<$ 10\%) and 
a flat PPA swing.  Pulsar B1530+27's postcursor (bottom left) exhibits somewhat 
different polarization characteristics.  Its linear polarization is smaller ($<$ 30\%) 
but with large circular polarization ($>$ 70\%).  Its PPA swing, however, is 
relatively flat across the feature.  The precursor in pulsar B1822--09 (bottom center) 
is characterized by high linear polarization ($>$ 80\%) and minimal circular 
polarization ($<$ 10\%).  The PPA is relatively flat compared to the main pulse but 
shows a change of about 30$\degr$ across the precursor. 
\label{fig_pol}}
\end{figure*}

\begin{figure*}
\includegraphics[angle=0,scale=1.33]{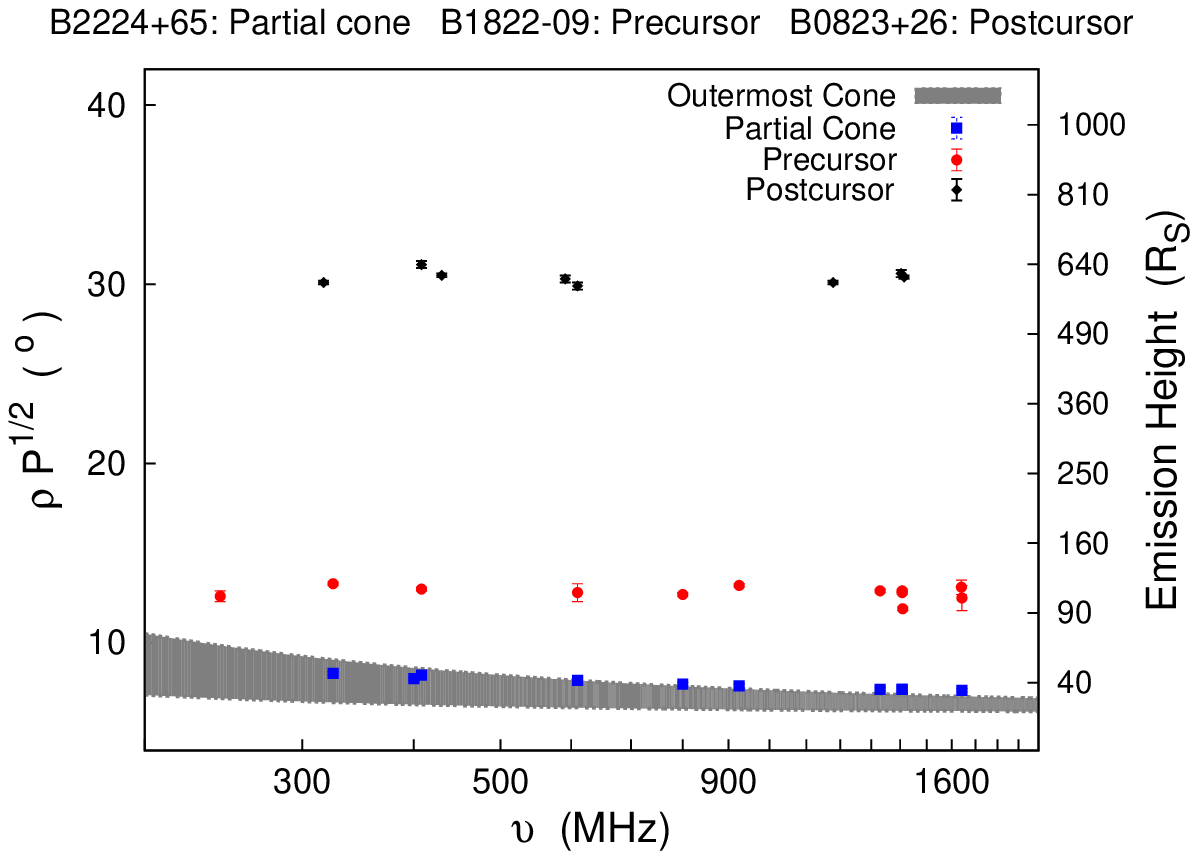}
\caption{Evolution of the beam opening-angle radius with frequency for the 
precursor in B1822--09, postcursor in B0823+26 and the partial cone in B2224+65. 
The shaded region represents the outermost cone as shown in eq.\ref{eqn_conemit} 
($\varrho$ = 1.3).  The boundaries of the shaded region signify the maximum and minimum 
errors in determining the conal boundary.  The partial cone component in B2224+65 exhibits 
decreasing separation from the pulse center with increasing frequency and has been 
identified as an extremely outer region of the beam and not a postcursor component.  
The precursor and postcursor components are not consistent with the core-cone 
morphology, showing no frequency evolution of separation and lie outside the conal 
beam opening angle.  In the right hand side we have shown the emission heights for the 
beam radius corresponding to the last open magnetic field lines.  The PPC components 
originate much higher up along the open field lines compared to the main pulse 
emission.
\label{fig_rad}}
\end{figure*}

\section{\large Discussion}
\subsection{Properties of PPC emission}
\noindent
The PPC emission is asymmetric, without any counterpart at the opposite side of 
the pulse center, and thereby resembles a partial cone.  It is generally connected 
to the main pulse with a low level emission bridge, and exhibits a high degree of 
linear polarization with a relatively flat PPA traverse.  Our analysis, carried out 
assuming that all emission components originate along the open dipolar magnetic 
field lines, reveal the PPCs to be distinct from the outer conal emission, and we 
highlight some of their characteristic features.  

\noindent
{\bf Separation of components:}  PPC emission features, much like interpulses or 
inner conal component pairs, show little or no significant spectral change in their 
spacing from the main pulse region.  This property serves to differentiate PPC 
features from outer conal component pairs, which regularly show the widening 
with wavelength known as ``radius-to-frequency mapping''.  
Interpulse and main pulse emission are expected to originate from opposite magnetic 
poles, with the interpulse visible only for pulsars with an orthogonal geometry.  PPC 
emission is then quite unlike that from either polar flux tube, and it seems likely 
that it is emitted at some different location within the pulsar magnetosphere.  However, 
given that the PPC-main pulse separations are usually much less than 180$\degr$,  
the PPC emission is likely to arise from the same magnetic pole as the main 
pulse.  

\noindent
{\bf Beam opening angle:}  Outer conal emission, to the extent that it originates 
near the periphery of the polar flux tube, represents the extremity of main pulse 
emission in terms of both opening angle and altitude.  The beam opening angle 
changes with frequency and is estimated using the geometry of the pulsar and 
the longitudinal extent of the component in question (see eq. \ref{eqn_beamwid}).  
The upper limit for the width of a beam of outer conal emission turns out to 
be around 7.5$\degr$ at 1 GHz (eq. \ref{eqn_conemit}, here the beam radius is 
normalized by P$^{0.5}$). The partial cone in pulsar B2224+65 has a beam 
opening angle of 7.6$\degr$ and is consistent with the conal width. The precursor 
emission in B0943+10, B0950+08, B1322+83, B1524--39, and B1822--09 has 
beam opening angles between 10--15$\degr$, while the postcursor components 
in B0823+26, B1530+27 and B2217+47 have beam opening angles around 
30$\degr$.\footnote{The beam opening angle of the postcursor in B1929+10 is 
around 50\degr, but this value is likely to be lower as we have used the most 
extreme case of orthogonal geometry for calculations; see section 3 for details.
Similarly, the large beam opening angle of the postcursor in B0940+16 may very 
well be due to errors in determining the pulsar geometry.} 
The beam opening angles based on the separation between the PPC and 
main pulse are much greater than the conventional outer conal emission and 
form the distinguishing feature  between them.  It is interesting to note that the 
beam radii of the precursor and postcursor emission are different, with the latter 
values being greater by a factor of two.  We have too few examples of PPC 
emission to draw any more general conclusion regarding their differences.\\

\noindent
{\bf Emission Heights:} The radio emission in pulsars is expected to originate 
along the open dipolar magnetic field lines. The outer conal emission lies in 
the outermost regions of the emission beam and its beam opening angle 
provides an estimate of emission height. The emission heights using the minimum 
and maximum error values of outermost beam opening angles specified in 
eq.\ref{eqn_conemit} and eq.\ref{eqn_hgeo} turns out to be 25--35 $R_S$ at 
1 GHz for a 1 second pulsar. Using the beam opening angle the emission height for 
the precursor emission is calculated to be around 100 $R_S$ and the postcursor emission 
is more than 500 $R_S$ (see Figure~\ref{fig_rad}). The above estimates should be 
taken as lower limits as we do not yet know the exact location of the PPC 
components within the emission beam. It is clear that the PPC emission has a 
location much higher up in the pulsar magnetosphere compared to the main pulse 
emission. The implications of the emission height on the underlying mechanism 
of emission are discussed in the next subsection.\\

\noindent
{\bf Polarization Characteristics:} The PPC components of B0823+26, B0943+10, B1322+83,
B1530+27 and B1822--09 with their polarization properties are shown in figure~\ref{fig_pol}.
The polarization characteristics of the PPC emission in the well known pulsars B0950+08 
and B1929+10 are shown in \citet{mcl04} (see their figure 4 and 5).  The PPC emission is 
associated with a high degree of linear polarization greater than 70 percent in most cases. 
The circular polarization is much lower and is usually less than 20 percent. The notable 
exception being B1530+27 where the postcursor emission shows relatively low linear polarization, 
less than 30 percent, and high circular polarization, greater than 70 percent. The PPA remains 
relatively flat across the PPC components compared to the swing across the main pulse. A closer 
look reveals the PPA to be not completely flat but shows some gentle swing with the change being 
as much as 20--40$\degr$ across some PPC components (see Figure~\ref{fig_pol}). The high degree 
of linear polarization as well as the absence of orthogonal modes in the PPC emission indicate 
the likely source of emission to be a pure polarization mode.  

For pulsar B0823+26 we are able to securely identify the alignment of the 
dominant postcursor polarization mode.  As summarized in \citet{ran07},
\citet{mor79} measured the absolute PPA at the peak of the star's main pulse 
as +48(3)\degr counter-clockwise from North.  Its proper motion direction was determined 
by \citet{lyn82} as +146(1)\degr.  Assuming that the former reflects the ``fiducial'' 
orientation of the projected magnetic field at the axis longitude, then the 
latter minus the former indicates an orthogonal 98(1)\degr~relation between the 
projected field orientation and the polarization of the radiation, implying that 
this radiation represents the extraordinary (X) propagation mode.  Further, 
reference to the \citet{eve01} discussion on this pulsar shows clearly that the 
postcursor polarization represents the same mode (X) as the peak of the main pulse.

\subsection{PPC emission in terms of emission mechanisms}

\begin{table*}
\caption{Conditions for coherent curvature radiation in PPC emission. 
\label{tab_coh}}
\begin{center}
\leavevmode
\begin{tabular}{lcccccccccc}
\hline\hline
~~Pulsar  &   P   &  $\dot{P}_{-15}$ & B$_S$ & $\dot{E}$ & $r_{PC}$ & $\nu_p$ & eq.(\ref{eq_growth}) & $\gamma_c$ & $\nu_c$ & $\gamma_c^{500}$\\
          & s &10$^{-15}$ s s$^{\tiny -1}$& 10$^{12}$ G & 10$^{31}$ erg s$^{-1}$ &$R_S$&MHz& $\gamma_p$=200& &MHz& \\
\hline
  B0823+26  & 0.531 & ~1.71 & 0.96 & ~45.2 & 600 & ~21.6 & 0.011 & ~10~ & 0.001 & 750 \\
  B0943+10  & 1.098 & ~3.49 & 1.98 & ~10.4 & 130 & 213.1 & 0.113 & ~45~ & 0.559 & 450 \\
  B0950+08  & 0.253 & ~0.23 & 0.24 & ~56~~ & ~70 & 392.8 & 0.209 & ~70~ & 1.950 & 350 \\
  B1322+83  & 0.670 & ~0.57 & 0.62 & ~~7.4 & ~75 & 349.8 & 0.185 & ~65~ & 2.610 & 375 \\
  B1524--39 & 2.418 & 19.07 & 6.87 & ~~5.3 & 110 & 343.7 & 0.182 & ~65~ & 1.717 & 425 \\
  B1530+27  & 1.125 & ~0.78 & 0.95 & ~~2.2 & 550 & ~16.7 & 0.009 & ~10~ & 0.0008& 725 \\
  B1822--09 & 0.769 & 52.50 & 6.43 & 456~~ & 110 & 589.5 & 0.311 & ~90~ & 5.052 & 425 \\
  B2217+47  & 0.538 & ~2.77 & 1.23 & ~69.9 & 800 & ~15.7 & 0.008 & ~10~ & 0.001 & 825 \\
\hline
\end{tabular}
\tablenotetext{~}{$\nu_p$ = $\omega_p$/2$\pi$ and $\nu_c$ = $\omega_c$/2$\pi$. The condition for 
coherent radio emission requires eq.(\ref{eq_growth}) and $\nu_c <$ 2$\sqrt{\gamma_p} \nu_p$. The 
conditions are not satisfied in all of the cases, see section 4.2 for details.} 
\end{center}
\end{table*}

\noindent
The absence of frequency evolution of PPC emission leads to two possibilities 
about its origin; either they emanate from a range of emission altitudes with
no discernible frequency evolution (like in regions of the magnetosphere where 
the magnetic field lines do not diverge significantly with emission height) or 
the emission mechanism has provisions for emitting different frequencies at 
similar locations in the magnetosphere. Based on a number of phenomenological 
properties of the main pulse radio emission, the most favourable mechanism 
appears to be the soliton induced curvature radiation \citep{mel00,gil04,mit09}.
Here we will discuss whether the same mechanism can be applied to the PPC 
emission as well. 

The soliton model requires the presence of a nonstationary flow of the plasma 
along the open magnetic field lines. The presence of an inner vacuum gap above 
the magnetic poles acts as a source of relativistic plasma \citep{rud75,gil03} 
comprised of singly charged primary particles (characterised by Lorentz 
factor $\gamma_b \sim$ 10$^6$) and secondary pair plasma (with Lorentz
factors $\gamma_p \sim$ 10--1000). The nonstationary flow coupled with the 
spread in particle energies results in overlapping clouds of secondary plasma.  
This gives rise to the two-stream instability which generates electrostatic 
Langmuir waves \citep{ase98}. The modulational instability of the Langmuir waves 
leads to the formation of relativistic charged solitons which excite coherent 
curvature radiation in the ambient plasma. The condition for coherent emission 
requires the wavelength to be greater than the intrinsic soliton size \citep{mel00}, 
\ie
\begin{equation}
\omega < 2\sqrt{\gamma_p} \omega_p,
\label{eq_freq}
\end{equation}
where $\omega_p$ = $(4\pi e^2 n/m_e)^{1/2}$ is the plasma frequency; 
$n$ = $\chi n_{GJ}$, with $\chi$ $\sim$ 10$^4$. 

The characteristic frequency of curvature radiation is given as
\begin{equation}
\omega_c = \frac{3}{2}\gamma_p^3 c/R_C,
\label{eq_cur}
\end{equation}
where $R_C$ is the curvature radius of the magnetic field lines.  The emission 
occurs at heights of 50$R_S$ where the radius of curvature of dipolar magnetic 
field lines can be approximated as $R_C \sim 10~r$, here $r$ being the
emission height.  The particle energies required for emission at these heights
are $\gamma_c$ $\sim$ 200. The growth of the Langmuir waves is determined by 
the complex frequency ($\Gamma$) where the instability can grow if the growth 
time (1/$\Gamma$) is much less than the characteristic time scale of the plasma 
cloud to overlap and interact (1/$\Gamma$ $\ll \phi_o$/c; here $\phi_o \sim$ 
10$^5$ cm, is the length scale of secondary plasma cloud). The inequality is 
expressed in terms of typical pulsar parameters as \citep{ase98}:
\begin{equation}
\small
(\gamma_p /100)^{-1.5} (r/50R_S)^{-1.5}(\dot{P}_{-15}/P)^{1/4} \gg 0.1,
\label{eq_growth}
\end{equation}
where $P$ is the pulsar period, $R_S$ the radius of neutron star and 
$\dot{P}_{-15}$ the period derivative in units of 10$^{-15}$ s s$^{\tiny -1}$. 

We have applied the soliton induced curvature radiation mechanism to the PPC 
emission in eight pulsars and the results are outlined in Table \ref{tab_coh}
along with the basic pulsar properties period, period derivative, surface 
magnetic field and spin down energies. The emission heights are determined from 
the radii of the beam opening angles (using eq.~\ref{eqn_hgeo}) and vary from 
70--800 $R_S$, see column 6~in the table, which as we mentioned earlier are lower 
limits. We undertake the following calculations to determine the viability of the 
emission mechanism for PPC components.\\
1. The values of the plasma frequency are determined in column 7 for each pulsar at 
the emission height of the PPC components. The plasma frequencies are quite low which 
already makes it unlikely for the mechanism to emit higher frequencies ($>$ 1 GHz) where 
PPC emission is observed (see eq.~\ref{eq_freq}).\\
2. We checked whether the particles responsible for the main pulse emission
with energies $\gamma_p \sim$ 200 can give rise to the detected PPC components.
The condition in eq.(\ref{eq_growth}) is determined for the $\gamma_p$ = 200 particles 
(Table \ref{tab_coh}, column 8) and turns out to be low for optimal growth. These clearly 
imply that the $\gamma_p \sim$ 200 particles in the secondary plasma are incapable of 
developing the necessary instability growth to generate coherent emission at these 
emission heights.\\
3. We next look at the particles in the secondary plasma which can give rise to 
optimal growth. We assumed the condition in eq.(\ref{eq_growth}) to be unity and 
determined the particle energies which satisfy the growth condition at the PPC 
emission heights. The particle energies turn out to be be very low, $\gamma_c$ is 
between 10 to 100 as shown in column 9. We have determined the frequency of curvature 
radiation ($\nu_c$) at these emission heights using $\gamma_c$ for each of the pulsars,
and the emission frequency is very low ranging from a few kilohertz to a few 
megahertz---much below the observing frequency as shown in column 10 of Table 
\ref{tab_coh}.  This clearly indicates that the particles in the secondary 
plasma which are suitable for growth of instability would emit at much below 
observing frequencies.\\
4. Finally, we have calculated the particle energies which are capable of
curvature radiation in the observable RF range.  We took the radiation frequency 
as 500 MHz and using eq.(\ref{eq_cur}) calculated $\gamma_c^{500}$ at the
heights of PPC emission.  The Lorentz factors ranged from 350 to 825 as shown
in the last column.  Lorentz factors between 350 and 825 would violate 
eq.(\ref{eq_growth}) which will not lead to coherent emission.

We have demonstrated that the soliton induced coherent curvature radiation which 
is responsible for the main pulse emission is unlikely to emit PPC components. 
Note that in the case of PSR B0823+26, we have explored a scenario where the 
PPA below the PPC as well as the main pulse can be associated with the X plasma mode, 
which in turn argues for a mechanism similar to curvature radiaiton. This issue needs 
further understanding which however is beyond the scope of our present studies. 
 
There is another emission mechanism which develops in the outer magnetosphere 
near the light cylinder which can give rise to coherent radio emission 
\citep{kaz91,lyu99}. The coherent emission can be generated when the naturally 
developing electromagnetic modes in the clouds of secondary plasma undergo 
negative absorption from the highly energetic primary particles via the 
cyclotron resonance instability. Recently \citet{bas13} have used this 
mechanism to explain the origin of off-pulse emission from long period pulsars.
However, the mechanism suffers from concerns of cutoff frequencies and fine 
tuning of plasma parameters similar to the curvature emission mechanism. This 
scenario would further require explanation of the pulsed nature of the PPC 
emission and detailed studies are needed to ascertain its viability as a 
possible emission mechanism.

\citet{pet08} has explained the precursor emission as a result of induced 
scattering of the main pulse emission into the background.  The scattered 
precursor is directed along the magnetic field and appears ahead of the main
pulse due to rotational aberration.  However, according to this mechanism the
precursor separation does not exceed 30$\degr$ in longitude for the scattering 
region to be within the magnetosphere. In our studies we have found precursors 
in pulsars B0943+10, B0950+08, B1322+83 and B1524--39 with longitudinal separation 
in excess of 30$\degr$. This study further does not address the presence of 
postcursors.

In this work we have demonstrated the precursor and postcursor emission in 
pulsars to be different from the canonical main pulse emission.  The locations 
of the PPC emission are higher up in the magnetosphere compared to the main 
pulse. It seems likely that the coherent curvature radiation mechanism for the 
main pulse is inadequate to explain their origin.  The number of pulsars with
precursor components stands at five while there are four pulsars with postcursor 
emission. We have also identified three potential candidates for PPC emission 
which we could not fully analyze due to the lack of high quality observations.  
Detailed studies are required to search for more PPC emission in the pulsar 
population and to further characterize their nature including identifying the 
location of the PPC emission within the magnetosphere and also looking for an 
emission mechanism applicable to them.

\noindent
{\bf Acknowledgments}: We thank the anonymous referee for useful comments. We would also like 
to thank Janusz Gil and Giorgi Melikidze for their comments which helped to improve the paper. 
This work was partially supported by the grant DEC-2013/09/B/ST9/02177 of the Polish National 
Science Centre. Portions of this work were carried out with support from US National Science 
Foundation grants 08-07691 and 09-68296.  Arecibo Observatory is operated by SRI International 
under a cooperative agreement with the NSF, and in alliance with Ana G. M\'endez-Universidad 
Metropolitana, and the Universities Space Research Association.  We thank the staff of the GMRT 
that made the observations possible. GMRT is run by the National Centre for Radio Astrophysics 
of the Tata Institute of Fundamental Research. This work made use of the NASA ADS astronomical 
data system.

\end{document}